\documentclass[aps,twocolumn,floatfix]{revtex4}

\usepackage{graphicx}
\usepackage{color}

\begin{document}

\bibliographystyle{apsrev}

\title{Forward Modeling of Space-borne Gravitational Wave Detectors}
\author{\surname{Louis} J. Rubbo}
\author{\surname{Neil} J. Cornish}
\author{\surname{Olivier} Poujade}

\affiliation{Department of Physics, 
  Montana State University, Bozeman, MT 59717}
\date{\today}

\begin{abstract}
Planning is underway for several space-borne gravitational wave
observatories to be built in the next ten to twenty years.  Realistic
and efficient forward modeling will play a key role in the design and
operation of these observatories. Space-borne interferometric
gravitational wave detectors operate very differently from their
ground based counterparts. Complex orbital motion, virtual
interferometry, and finite size effects complicate the description of
space-based systems, while nonlinear control systems complicate the
description of ground based systems. Here we explore the forward
modeling of space-based gravitational wave detectors and introduce
an adiabatic approximation to the detector response that significantly
extends the range of the standard low frequency approximation. The
adiabatic approximation will aid in the development of data analysis
techniques, and improve the modeling of astrophysical parameter
extraction.
\end{abstract}

 \maketitle

\section{Introduction}

Gravitational wave astronomy can be broadly divided into high and low
frequency bands, with the dividing line near one Hertz.  Seismic and
gravity gradient noise prevent ground based detectors from exploring
the low frequency portion of the spectrum, making this source-rich
region the sole preserve of space-based observatories.

Ground and space-based interferometric gravitational wave detectors
operate according to the same general principles, but differ
in their implementation. Ground-based detectors, such as the
Laser Interferometer Gravitational Wave Observatory (LIGO)~\cite{Aetal92},
operate in the low frequency limit, where the
wavelength of the gravitational waves is considerably larger than the
size of the detector, and most sources are only in-band for a fraction
of a second. These considerations simplify the description of the
detector response, which may be well approximated by a quadrupole
antenna moving at constant velocity with respect to the gravitational
wave source. However, ground-based interferometers employ quasi-fixed
rather than freely moving test masses, and the output of the detector
is given by the response of the control loop used to keep the
interferometer on a dark fringe. This complicates forward modeling
efforts for ground-based detectors~\cite{Setal99} as it makes the
detector response non-linear.  The situation with space-borne detectors
is completely the opposite.  Space-based detectors, such as the
proposed Laser Interferometer Space Antenna (LISA)~\cite{Vetal02},
will be able to detect gravitational waves with wavelengths that range
from many times larger than the interferometer to many times smaller,
and most sources will be in-band for months or years, so that the
detector's orbital motion will impart amplitude, frequency, and phase
modulations. These effects give rise to a complicated, time dependent
detector response function~\cite{CR03}.  Space-borne detectors
typically have large arm-lengths ($5 \times 10^9$ m for LISA) that
vary with time, which prevents them from operating as traditional
interferometers.  Instead, the interferometer signals are produce in
software from phase differences measured in the detector using a
procedure know as Time Delay Interferometry (TDI)~\cite{TDI}.  Despite
these complications, the detector response remains linear, which greatly
simplifies forward modeling efforts.

Forward modeling plays a key role in the design of any new scientific
instrument, and is especially important when the instrument is the first of
its kind.  Work is now underway to produce an end-to-end model of the
LISA observatory~\cite{M03}.  Key ingredients include accurate
modeling of the spacecraft orbits and photon trajectories (this
includes the effects of gravitational waves), realistic simulations of
the time delay interferometry used to cancel laser phase noise, and
experimental characterization of the various noise contributions. A
good end-to-end model can help to make design trade-offs, and to avoid
costly mistakes. Forward modeling can also be used to develop and test
data analysis strategies. While we focus our attention on LISA, our
forward model can be used to study other proposals for space-borne
gravitational wave detectors, such as the Big Bang Observatory~\cite{Petal03}.

Work on various elements of the LISA end-to-end model have been under
development for some time. Modeling of the detector response has its
roots in the Doppler tracking of spacecraft~\cite{EW75}. Results were
initially derived for a static array with equal arm-lengths~\cite{S97,LHH00}.
Following the discovery of Time Delay Interferometry~\cite{TDI},
these results were extended to a static array with unequal
arm-lengths~\cite{TDI,TDI2,LHH02}. The orbital motion of the array was
first incorporated in the low frequency limit~\cite{C98}, and later
extended to the full detector response~\cite{CR03}. With the full
response function in hand, we have developed an open source software
package called {\em The LISA Simulator}~\cite{TLS} that takes as its
input an arbitrary gravitational wave and returns as its output the
simulated response of the LISA observatory. The main purpose of {\em
The LISA Simulator} is to aid in the development of data analysis
tools~\cite{C98, MH02, CL03a, CL03b}, but its modular design allows it
to be extended into a full end-to-end model. For example, the static
modeling~\cite{AEI03} of the TDI implementation could be incorporated
into {\em The LISA Simulator}, as could more realistic spacecraft
orbits and experimentally determined noise spectra.

The value of a realistic end-to-end model has already become apparent
with the discovery of flaws in the initial TDI scheme caused by the
rotation of the array~\cite{S03}, time dependence of the
arm-lengths~\cite{CH03}, and problems with clock synchronization
in a moving array~\cite{TEA03}. These difficulties require modification
of the TDI variables~\cite{S03,CH03,STEA03} and/or changes in the
mission design.

On the other hand, a highly realistic end-to-end simulation
necessarily consumes a great deal of computer resources, and delivers
a fidelity that exceeds the requirements of many data analysis
efforts. Indeed, when searching a large parameter space, fidelity must
be sacrificed in favor of speed. To this end we have developed an
approximation to the full LISA response that extends the low frequency
approximation by two decades. The motion of the array is
stroboscopically rendered into a sequence of stationary states,
yielding an adiabatic approximation to the full response. The
adiabatic approximation allows us to write down a simple analytic
expression for the response function in a mixed time/frequency
representation. For sources with a few dominant harmonics, such as low
eccentricity, low spin binary systems at second post-Newtonian oder,
the adiabatic approximation provides a fast and accurate method for
calculating the LISA response.

The outline of this paper is as follows: In Sec.~\ref{Ssbd} we
describe the orbits of the interferometer constellation and describe
how various effects enter into the detector response. In
Sec.~\ref{Sdra} we review the expression for the complete response of
a space-borne detector. (An alternative derivation of
the full response is given in Appendix B). In Sec.~\ref{Sdrn} we show some
applications of the general formalism using {\em The LISA
Simulator}. In Sec.~\ref{Slfa} we explore the limitations of the low
frequency approximation, and in Sec.~\ref{Sraa} we introduce the
adiabatic approximation and demonstrate its utility. We finish with an
application, using the adiabatic approximation to determine when LISA
can detect the time evolution of a binary system. We work in natural
units with $G=c=h=1$, but report all frequencies in Hertz.

\section{Space-borne detectors}\label{Ssbd}

\subsection{Orbital effects} \label{Ssbd-oe}

\begin{figure}[!tb]
\begin{center}
\input{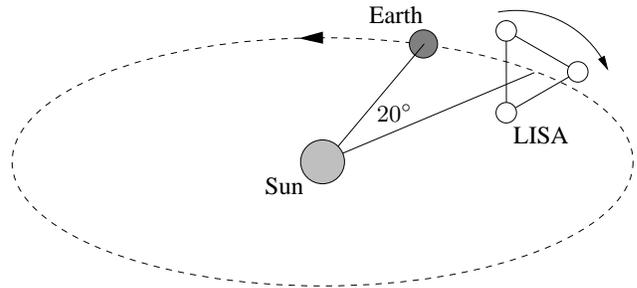}
\end{center}
\caption{The LISA mission configuration.  The dashed line represents
  the orbit of the guiding center, which has a radius of 1 AU.}
\label{LISA}
\end{figure}

The current design of the LISA mission calls for three identical
spacecraft flying in an equilateral triangular formation about the
Sun.  The center of mass for the constellation, known as the guiding
center, is in a circular orbit at 1 AU and $20^\circ$ behind the
Earth.  In addition to the guiding center motion, the formation will
cartwheel in a retrograde motion with a one year period (see
Fig. \ref{LISA}).  The detector motion introduces amplitude (AM),
frequency (FM), and phase modulations (PM) into the gravitational wave
signals \cite{C98, CL03b}.  The amplitude modulation is caused by the
antenna pattern being swept across the sky. The phase modulation
occurs when the differing responses to the two gravitational wave
polarizations are combined together. The frequency
(Doppler) modulation is due to the motion of the detector relative to
the source.  Since both the orbital and cartwheel motion have a period
of one year, these modulations will show up as sidebands in the power
spectrum separated from the instantaneous carrier frequency by integer
values of the modulation frequency, $f_m = 1/$yr.

To describe the coordinates of the detector we work in a heliocentric,
ecliptic coordinate system.  In this system the Sun is placed at the
origin, the $x$-axis points in the direction of the vernal equinox,
the $z$-axis is parallel to the orbital angular momentum vector of the
Earth, and the $y$-axis is placed in the ecliptic to complete the
right handed coordinate system.  Ignoring the influence from other
solar system bodies, the individual LISA spacecraft will follow
independent Keplerian orbits.  The triangular formation comes about
through the judicious selection of initial conditions.  In Appendix A
we derive the spacecraft positions as a function of time. To second
order in the eccentricity, the Cartesian coordinates of the spacecraft
are given by
\begin{eqnarray} \label{keporb}
x(t) &=& R \cos(\alpha) + \frac{1}{2} e R\Big( \cos(2\alpha-\beta) -
3\cos(\beta) \Big) \nonumber\\
&& + \frac{1}{8} e^2 R\Big( 3\cos(3\alpha-2\beta) - 10\cos(\alpha) 
\nonumber\\
&& - 5\cos(\alpha-2\beta) \Big) \nonumber\\
y(t) &=& R \sin(\alpha) + \frac{1}{2} e R\Big( \sin(2\alpha-\beta) -
3\sin(\beta) \Big) \nonumber\\
&& + \frac{1}{8} e^2 R\Big( 3\sin(3\alpha-2\beta) - 10\sin(\alpha)
 \nonumber\\
&& + 5\sin(\alpha-2\beta) \Big) \nonumber\\
z(t) &=& -\sqrt{3} e R\cos(\alpha-\beta) \nonumber\\
&&+ \sqrt{3} e^2 R \Big( \cos^2(\alpha-\beta) 
+ 2\sin^2(\alpha-\beta) \Big) \; .
\end{eqnarray}
In the above $R = 1$ AU is the radial distance to the guiding center,
$e$ is the eccentricity, $\alpha = 2\pi f_m t + \kappa$ is the orbital
phase of the guiding center, and $\beta = 2\pi n/3 + \lambda$
($n=0,1,2$) is the relative phase of the spacecraft within the
constellation.  The parameters $\kappa$ and $\lambda$ give the initial
ecliptic longitude and orientation of the constellation.

Using the above coordinates the instantaneous separations between
spacecraft are found to be
\begin{eqnarray} \label{ovar}
L_{12}(t) &=& L \bigg( 1 + \frac{e}{32} \bigg[ 15 \sin \Big( \alpha -
\lambda + \frac{\pi}{6} \Big) \nonumber\\
&& - \cos \Big(3(\alpha-\lambda)\Big) \bigg] \bigg) \nonumber\\
L_{13}(t) &=& L \bigg( 1 + \frac{e}{32} \bigg[ -15 \sin \Big( \alpha -
\lambda - \frac{\pi}{6} \Big) \nonumber\\
&& - \cos \Big(3(\alpha-\lambda)\Big) \bigg] \bigg) \nonumber\\
L_{23}(t) &=& L \bigg( 1 - \frac{e}{32} \bigg[ 15 \cos( \alpha -
\lambda ) \nonumber\\
&& + \cos \Big(3(\alpha-\lambda)\Big) \bigg] \bigg) \, ,
\end{eqnarray}
with $L = 2 \sqrt{3} R e$.  From this it is seen that to linear order
in the eccentricity the detector arms are rigid.  By setting the
mean arm-length equal to those of the LISA baseline, $L = 5 \times
10^9$ m, the spacecraft orbits are found to have an eccentricity of
$e=0.00965$, which indicates that the second order effects are down by a
factor of 100 relative to leading order.

\subsection{Gravitational Wave Description}

An arbitrary gravitational wave traveling in the $\hat{k}$ direction
can be written as the linear sum of two independent polarization
states,
\begin{equation}\label{wave}
{\bf h}(\xi) = h_+(\xi) {\mbox{\boldmath$\epsilon$}}^+
 + h_\times(\xi) {\mbox{\boldmath$\epsilon$}}^\times \,,
\end{equation}
where the wave variable $\xi = t - \hat{k} \cdot {\bf x}$ gives the
surfaces of constant phase. The polarization tensors are given by
\begin{eqnarray}
{\mbox{\boldmath$\epsilon$}}^+ &=& \cos(2\psi) {\bf e}^+ - \sin(2\psi)
  {\bf e}^\times \nonumber\\
{\mbox{\boldmath$\epsilon$}}^\times &=& \sin(2\psi) {\bf e}^+ +
\cos(2\psi) {\bf e}^\times \, ,
\end{eqnarray}
where $\psi$ is the principle polarization angle and the
basis tensors ${\bf e}^+$ and ${\bf e}^\times$ are expressed in terms
of two orthogonal unit vectors,
\begin{eqnarray}
{\bf e}^+ &=& \hat{u} \otimes \hat{u} - \hat{v} \otimes \hat{v}
\nonumber\\
{\bf e}^\times &=& \hat{u} \otimes \hat{v} + \hat{v} \otimes \hat{u} \,.
\end{eqnarray}
These vectors, along with the propagation direction of the
gravitational wave, form an orthonormal triad, which may be expressed
as a function of the source location on the celestial sphere
$(\theta,\phi)$,
\begin{eqnarray}
\hat{u} &=& \cos\theta\cos\phi\,\hat{x} + \cos\theta\sin\phi\,\hat{y} -
\sin\theta\,\hat{z} \nonumber\\
\hat{v} &=& \sin\phi\,\hat{x} - \cos\phi\,\hat{y} \nonumber\\
\hat{k} &=& -\sin\theta\cos\phi\,\hat{x} - \sin\theta\sin\phi\,\hat{y} -
\cos\theta\,\hat{z} \,.
\end{eqnarray}

The above basis set is defined with respect to the barycenter reference
frame. For a binary system - the standard gravitational wave source
in the LISA band - it is natural to introduce another basis that
is aligned with principle polarization axes, $\hat{p}$ and $\hat{q}$,
of the gravitational radiation.  The orientation of the principle
directions is chosen such that there is a $\pi/2$ phase delay between
the two polarization states.  The connection between the two basis
sets is a rotation by the principle polarization angle $\psi$ about
the shared propagation direction $\hat{k}$.

We model the gravitational waves from a binary system according to
\begin{equation}\label{harmony}
h_{+,\times}(\xi) = \sum_{n} h^{(n)}_{+,\times} e^{ i n\Psi(\xi)}
\end{equation}
where $\Psi(\xi)$ is the orbital phase. The instantaneous frequency of
the $n^{\rm th}$ gravitational wave harmonic is given by
\begin{equation}
f_{n}(\xi) = \frac{n}{\pi} \frac{\partial \Psi}{\partial t} \, .
\end{equation}
Unless the binary is highly eccentric or highly relativistic, the
dominant emission will be quadrupolar, with frequency
$f(\xi)=f_2(\xi)$, and will be well described by the restricted
post-Newtonian approximation:
\begin{eqnarray}
&&h_+(\xi) = \frac{2 {\cal M}(\pi f(\xi))^{2/3}}
{D_L}\left(1 + \cos^2 \iota \right)\cos 2\Psi(\xi) \nonumber \\
&&h_\times(\xi) = -\frac{4{\cal M}(\pi f(\xi))^{2/3}}
{D_L}\cos \iota \sin 2\Psi(\xi) \,.
\end{eqnarray}
Here ${\cal M}$ is the chirp mass, $D_L$ is the luminosity distance
and $\iota$ is the inclination of the binary to the line of sight.
Higher post-Newtonian corrections, eccentricity of the orbit, and spin
effects will introduce additional harmonics.

\section{Detector response: analytical} \label{Sdra}

For two spatially separated test particles in free fall, the effect of
a passing gravitational wave is to cause the proper distance between
the masses to vary as a function of time. Finding the detector
response reduces to solving for the appropriate timelike and null
geodesics in the spacetime with the metric
\begin{equation} \label{combo}
ds^2 = -(1+2\phi) dt^2 +(1-2\phi)(dx^2 +dy^2 +dz^2) + h_{ij} dx^i dx^j
\, .
\end{equation}
In the above equation $\phi$ denotes the Newtonian potential set up by
various bodies in the Solar system and $h_{ij}$ denotes the
time-varying metric perturbation due to gravitational waves described
in the previous section. The relevant geodesics are those of the two
spacecraft, $\vec{x}_1(\tau_1)$, $\vec{x}_2(\tau_2)$, and the photons
sent from spacecraft 1 to 2, $\vec{x}_\nu(\lambda)$. We need to find
the path taken by the photon that leaves spacecraft 1 at time $t_1$
and arrives at spacecraft 2 at time $t_2$, which amounts to a classic
pursuit problem in curved spacetime.  The calculation must take into
account a host of factors, some due to the Newtonian potential, and
some due to the gravitational wave. During the time taken for the
photon to travel between the spacecraft, both effects are small and
can be treated independently.

The Newtonian potential leads to a variety of effects, such as a
Shapiro time delay $\Delta L / L \sim M_\odot /R$, gravitational
redshift $\Delta \nu/\nu \sim M_\odot L /R^2$, deflection of light
$\Delta \theta \sim M_\odot L /R^2$, and tidal flexing $\Delta L /L
\sim M_\odot L^2 /R^3$. Each of these effects is considerably larger
than any of the effects caused by the passage of the gravitational
wave, and they have to be subtracted before the gravitational wave
data analysis begins. The first step in the subtraction relies on us
being able to accurately model the orbital phase shifts using the
Solar System Ephemeris.  The second step in the subtraction employs a
high pass filter to remove the residuals from the orbital fit, which
occur at harmonics of the modulation frequency $f_m = 1/{\rm
yr}\simeq 3.2\times 10^{-8}$ Hz. The orbital effects, and the
procedure for their removal, should be included in the full end-to-end
model, even though they do not directly affect the response of the
detector to gravitational waves.

The effect of the gravitational wave on the phase shift can be found
by setting $\phi=0$ in Eq.~(\ref{combo}) and solving the geodesic
equation for the photons and the spacecraft in the metric perturbed by
the gravitational wave. There are two equivalent approaches for
finding the phase shift. The first approach is to find the Doppler
shift of the photon emitted by the first spacecraft and received by
the second. The Doppler shift is then integrated with respect to time to
give the phase shift. The Doppler derivation is given in Appendix
B. The second approach is to integrate along the photons trajectory to
find the path length variation caused by the gravitational wave~\cite{CR03}.
The expressions given in
Appendix B is valid to all orders in the spacecraft velocity $v$, and
to first order in the gravitational wave strain $h$. However, as we
explained in Ref.~\cite{CR03}, it is hard to justify keeping terms of
order $vh$ given that $v \sim 10^{-4}$. It would take a phenomenally
bright source, with a signal to noise ratio of $\sim 10^5$, for the
$vh$ cross terms to be noticeable. Working to leading order in $v$ and
$h$, the path length variation for a photon propagating from
spacecraft $i$ to spacecraft $j$ is given by
\begin{equation} \label{dell}
\delta\ell_{ij}(t) = \frac{1}{2} \, \frac{\hat{r}_{ij}(t) \otimes
  \hat{r}_{ij}(t)}{1-\hat{k} \cdot \hat{r}_{ij}(t)} :
  \int_{\xi_i}^{\xi_j} {\bf h}(\xi) d\xi \, ,
\end{equation}
where $\hat{r}_{ij}(t)$ points from test mass $i$ to mass $j$ and
${\bf h}(\xi)$ is the gravitational wave tensor in the
transverse-traceless gauge.  The colon here denotes a double
contraction, ${\bf a}:{\bf b} = a^{ij}b_{ij}$.

Applying Eq.~(\ref{dell}) to a pair of orbiting spacecraft requires
the careful evaluation of the $\hat{r}_{ij}(t)$ unit vectors. This
calculation is complicated by the motion of the spacecraft and the
finite speed of light.  For a photon emitted from spacecraft $i$ at
time $t_i$ and received at spacecraft $j$ at time $t_j$ the proper
evaluation of the unit vectors is
\begin{equation}
\hat{r}_{ij}(t_i) = \frac{{\bf x}_j(t_j) - {\bf
    x}_i(t_i)}{\ell_{ij}(t_i)} \, .
\end{equation}
The distance the photon travels between spacecraft is given implicitly
through the relationship
\begin{equation}
\ell_{ij}(t_i) = \big\| {\bf x}_j(t_i+\ell_{ij}(t_i)) - {\bf x}_i(t_i)
\big\| \, .
\end{equation}
Here we have used the fact that the reception time is the emission
time plus the time of flight for the photon.  We can numerically
estimate the magnitude of this point ahead effect by expanding the
photon propagation distance in a $v/c$ series:
\begin{equation}
\ell_{ij}(t_i) = L_{ij}(t_i)\Big(1 + \hat{r}_{ij}(t_i) \cdot {\bf
  v}_j(t_i) + {\mathcal O}(v^2)\Big)\, ,
\end{equation}
where ${\bf v}_j(t_i)$ is the velocity of spacecraft $j$ and
\begin{equation}
L_{ij}(t_i) = \big\| {\bf x}_j(t_i) - {\bf x}_i(t_i) \big\|
\end{equation}
is the instantaneous spacecraft separation.  For the LISA mission with
a mean arm-length of $5 \times 10^9$ m and spacecraft velocity $v
\approx 2\pi f_m R \approx 10^{-4}$, pointing ahead gives a first
order effect of approximately $10^5$ m.  For comparison, the orbital
effects given in Eq.~(\ref{ovar}) impart a variation in the photon
propagation distance of $10^7$ m.

An arbitrary gravitational wave can be decomposed into its frequency
components:
\begin{equation}
{\bf h}(\xi) = \int_{-\infty}^{\infty} {\bf \widetilde h}(f) e^{2\pi i
  f \xi} df\, .
\end{equation}
Such a decomposition allows us to rewrite Eq. (\ref{dell}) in the form
\begin{equation} \label{dell2}
\delta\ell_{ij}(t) = \ell_{ij}(t) \int_{-\infty}^{\infty} {\bf
  D}(f,t,\hat{k}):{\bf \widetilde h}(f) e^{2\pi i f \xi} df \, ,
\end{equation}
where the one-arm detector tensor is given by
\begin{equation}
{\bf D}(f,t,\hat{k}) = \frac{1}{2}\Big(\hat{r}_{ij}(t) \otimes
  \hat{r}_{ij}(t)\Big) {\mathcal T}(f,t,\hat{k}) \, ,
\end{equation}
and the transfer function is
\begin{eqnarray}
{\mathcal T}(f,t,\hat{k}) &=& \textrm{sinc} \left( \frac{f}{2
f^*_{ij}}\Big(1 - \hat{k} \cdot \hat{r}_{ij}(t)\Big) \right) \nonumber\\
&& \times \exp \left( i \frac{f}{2 f^*_{ij}}\Big(1 - \hat{k} \cdot
    \hat{r}_{ij}(t) \Big) \right) \, .
\end{eqnarray}
Here $f^*_{ij} = 1/(2\pi\ell_{ij})$ is the transfer frequency for the
$ij$-arm.  The transfer functions arise from the interaction of the
gravitational wave with the detector.  For gravitational radiation
whose frequency is greater than the transfer frequency the wave period
is less than the light propagation time between spacecraft, which
leads to a self-cancellation effect accounted for by the transfer
functions.  Below the transfer frequency the transfer functions
approach unity.  This leads to a natural division of the LISA
bandwidth into high and low frequency regions, which will be exploited
in a later section when we approximate the response of the detector.

The connection of Eq.~(\ref{dell}) to what is actually measured
depends on the design of the gravitational wave detector.
The current proposal for LISA is to have each spacecraft measure two
phases differences, one for each arm.  The phase difference,
$\Phi_{ij}(t_j)$, as measured on spacecraft $j$, is found by comparing
the phase of the received signal from spacecraft $i$ against the
outgoing signal's phase that is traveling back to spacecraft $i$.
Inherent in the phase difference measurements are both the
gravitational wave signal and noise contributions from laser phase
noise $C(t)$, shot noise $n^s(t)$, and acceleration noise $n^a(t)$:
\begin{eqnarray} \label{phi}
\Phi_{ij}(t_j) &=& C_{ji}(t_i) - C_{ij}(t_j) + 2 \pi \nu_0
\left(n_{ij}^s(t_j) - n_{ij}^a(t_j) \right. \nonumber\\
&& \left. + n_{ji}^a(t_i) + \delta_{ij}\ell(t_i)\right) \, .
\end{eqnarray}
Here the time $t_i$ is implicitly found through $t_i = t_j -
\ell_{ij}(t_i)$.  The subscripts on the noise components indicate the
directional dependence of that component: $C_{ij}$ is the laser phase
noise introduced by the laser on spacecraft $j$ that is pointed toward
spacecraft $i$, $n_{ij}^s$ is the shot noise in the photodetector on
spacecraft $j$ that is receiving a signal from spacecraft $i$, and
$n_{ij}^a$ is the projected acceleration noise from the accelerometer
on spacecraft $j$ in the direction of spacecraft $i$. The position
noise and path length variation are converted into a phase
difference by multiplying by the angular frequency of the laser, $2
\pi \nu_0$.

Once the six phase differences are measured and telemetered down, the
different interferometer signals can be synthesized.  For example, the
Michelson signal formed by using spacecraft 1 as the vertex craft is
\begin{equation}
S_1(t) = \Phi_{12}(t_{21}) + \Phi_{21}(t) - \Phi_{13}(t_{31}) -
\Phi_{31}(t) \, ,
\end{equation}
where $t_{21}$ and $t_{31}$ are found from
\begin{eqnarray}
t_{21} &=& t - \ell_{21}(t_{21}) \nonumber\\
t_{31} &=& t - \ell_{31}(t_{31}) \, .
\end{eqnarray}
However, due to the relatively large laser phase noise, the Michelson
signal will not be a viable option.  Instead a number of so called TDI
signals will be used~\cite{TDI}.  These
signals are built by combining time-delayed Michelson signals in such
a way as to reduce the overall laser phase noise down to a level that
will not overwhelm the detector's output.  A particular example of a
TDI variable is the $X$ signal~\cite{CH03}:
\begin{eqnarray} \label{xsig}
X(t) &=& \Phi_{12}(t_{21}) + \Phi_{21}(t) - \Phi_{13}(t_{31})
\nonumber\\ 
&& - \Phi_{31}(t) - \Phi_{12}(t_{21}') - \Phi_{21}(t_{13}) \nonumber\\
&& + \Phi_{13}(t_{31}') + \Phi_{31}(t_{12}) \, ,
\end{eqnarray}
where the new times $t_{12}$, $t_{13}$, $t_{21}'$, and $t_{31}'$ are
defined through the implicit relationships
\begin{eqnarray}
t_{12} &=& t_{21} - \ell_{12}(t_{12}) \nonumber\\
t_{13} &=& t_{31} - \ell_{13}(t_{13}) \nonumber\\
t_{21}' &=& t_{13} - \ell_{21}(t_{21}') \nonumber\\
t_{31}' &=& t_{12} - \ell_{31}(t_{31}') \, .
\end{eqnarray}
By permutations of the indices similar forms for the $Y$ and
$Z$-signals can be constructed.

By writing the response of the detector in a coordinate free manner we
are able to apply this formalism to an arbitrary space-based
mission. All that has to be changed are the spacecraft orbits.  It
should also be emphasized that the response is calculated entirely in
the time domain.  In later sections we develop approximations to the
full response by working in a hybrid time/frequency domain.  This
hybrid approach assumes extra information about the sources, which
allows us to develop explicit expressions for the detector response.

\section{Detector response: numerical} \label{Sdrn}

\subsection{Noiseless response}

As an application of the equations presented in the previous section,
we have simulated the response of the proposed LISA mission.  {\it The
LISA Simulator}~\cite{TLS} is designed to take an arbitrary
gravitational waveform and output the full response of the detector.
To apply the equations we have elected to work entirely in the
heliocentric, ecliptic coordinate system.  Therefore, all times are
evaluated in terms of Solar System Barycentric (SSB) time.  The
conversion to the detector time is through the standard relationship
$d\tau = \sqrt{1-v^2(t)}dt$, but since we only work to leading order
in $v$ the distinction is not made.  (In practice there will be
difficulties in synchronizing the clocks on the spacecraft~\cite{TEA03},
but they do not trouble the simulations.)

The positions of the spacecraft are calculated to second order in the
eccentricity, Eq.~(\ref{keporb}), which includes the leading order
flexing motion of the array. Tidal effects, and third order terms in
the eccentricity, are neglected for now.

\begin{figure}[!tb]
\begin{center}
\includegraphics[width=0.48\textwidth]{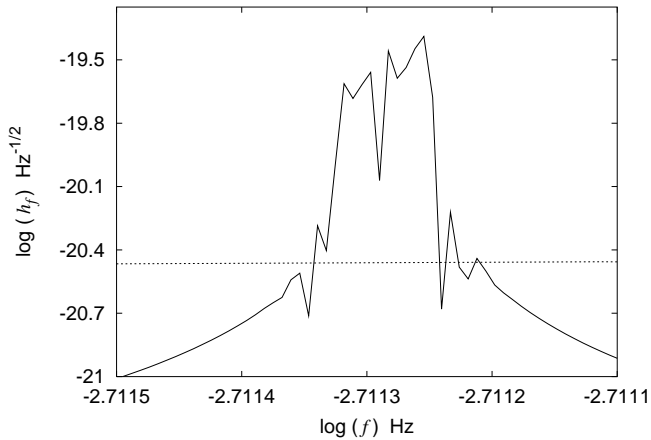}
\end{center}
\caption{The simulated $X$ strain spectral density of AM CVn,
  demonstrating the induced modulations caused by the motion of the
  detector about the Sun.  For reference, the dotted line is the
  average noise in this region of the spectrum.}
\label{fAMCVn}
\end{figure}

One of the guaranteed sources for the LISA mission is the cataclysmic
variable AM Canum Venaticorum.  This binary star system is comprised
of a low mass helium white dwarf that is transferring material to a
more massive white dwarf by way of Roche lobe overflow.  AM CVn's
orbital frequency of 0.972 mHz, and close proximity to the Earth
($\sim$100 pc) make it a good calibration binary for LISA.  Shown in
Fig.~\ref{fAMCVn} is the simulated response to AM CVn expressed as a
strain spectral density $h_f(f)$.  Note that the barycenter
gravitational wave signal will be approximately monochromatic,
however, the motion of LISA introduces modulations that cause the
signal to spread over a range of frequencies~\cite{CL03b}.

\begin{figure}[!tb]
\begin{center}
\includegraphics[width=0.48\textwidth]{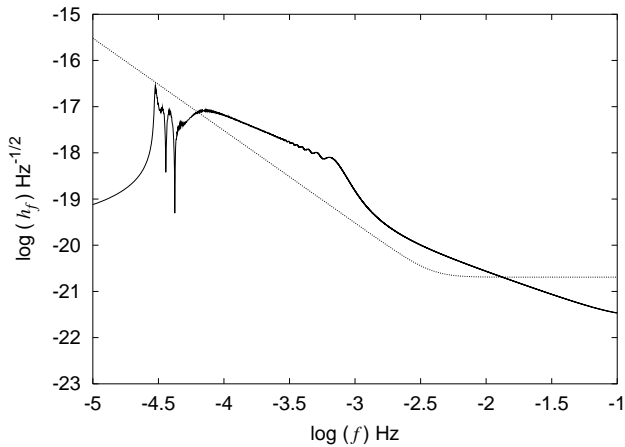}
\end{center}
\caption{The Michelson response of LISA to two $10^6
  \,\textrm{M}_\odot$ black holes coalescing at $z = 1$ ($D_L =
  6.63$ Gpc). The dotted line is the average Michelson noise in the
  detector.}
\label{BHmer}
\end{figure}

Another LISA source, but one whose event rate is poorly known, is the
merger of two super-massive black holes.  Figure~\ref{BHmer} shows the
simulated response of LISA to two $10^6 \,\textrm{M}_\odot$ black
holes coalescing at a redshift of $z=1$. The observation tracks the
final year before coalescence.

\subsection{Noise}

Laser phase noise, photon shot noise, and acceleration noise are
expected to be the dominant forms of noise in space-borne detectors.
As previously discussed, Time Delay Interferometry is used to
reduce the effects of the laser phase noise to a tolerable level. We
assume that the TDI signal processing is properly implemented, and
therefore neglect laser phase noise in our simulation.

\begin{figure}[!tb]
\begin{center}
\includegraphics[width=0.48\textwidth]{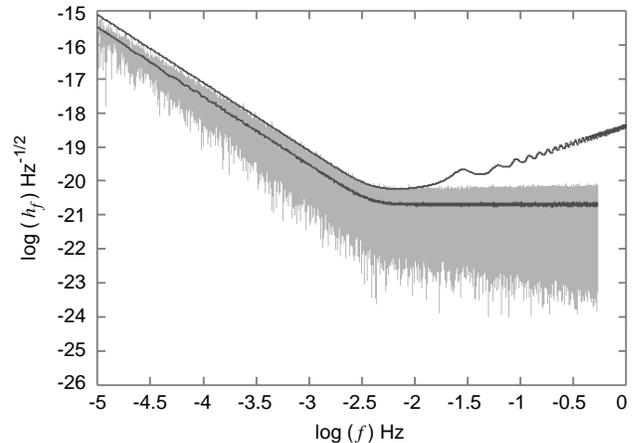}
\end{center}
\caption{A realization of the Michelson noise for LISA, expressed here
  as a strain spectral density.  The dark line within the noise is a
  128 bin rms value, while the rising curve is a standard LISA
  sensitivity curve.}
\label{Mnoise}
\end{figure}

The simulation of the noise is done in the time domain by drawing
random numbers at each time step from a Gaussian distribution with
unit variance and zero mean.  For the white photon noise we then
scale the random number by the shot noise spectral density
defined in Ref.~\cite{LPPA} ($S_{\rm ps} = 1.0 \times 10^{-22}
\,\textrm{m}^2 / \textrm{Hz}$).  For the colored acceleration noise we
begin by generating a white noise time series scaled by the
acceleration noise spectral density ($S_{\rm acc} = 9.0 \times 10^{-30}
\,\textrm{m}^2/\textrm{s}^4/\textrm{Hz}$), then integrate it twice to
arrive at a colored time series.  The integration introduces a $f^{-4}$
falloff in the power spectrum that is characteristic of acceleration
noise.  The results of this procedure for the Michelson signal are
shown in Fig.~\ref{Mnoise}.

Comparing this graph to a standard LISA sensitivity curve~\cite{SCG},
a number of differences are apparent. The most obvious one is the lack
of rise in the high frequency region. This is because the standard
sensitivity curve folds the average detector response into the noise
curve.  The {\it Sensitivity Curve Generator} includes
the all sky averaged and polarization averaged transfer function, which equals
$3/5$ at low frequencies and grows as $f^2$ above the transfer
frequency. A secondary difference is in the overall normalization, as
the {\it Sensitivity Curve Generator} scales the path length
variations by the interferometer mean arm-length of $L$, while we
scale the path length variations by the optical path length of $2L$.

\begin{figure}[!tb]
\begin{center}
\includegraphics[width=0.48\textwidth]{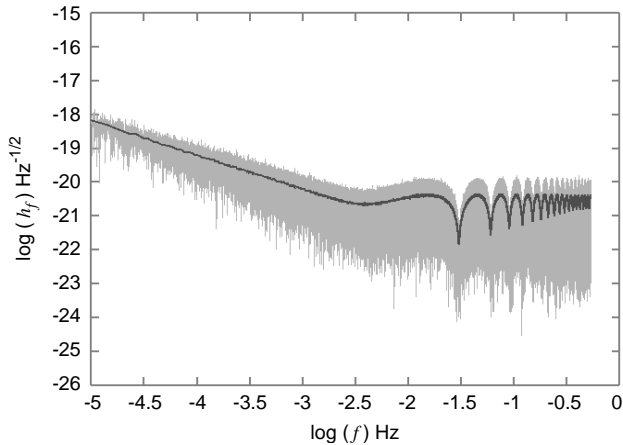}
\end{center}
\caption{A realization of the $X$ noise for LISA, expressed here as a
  strain spectral density.  The dark line is a 128 bin rms value for
  the noise.}
\label{Xnoise}
\end{figure}

\begin{figure}[!tb]
\begin{center}
\includegraphics[width=0.48\textwidth]{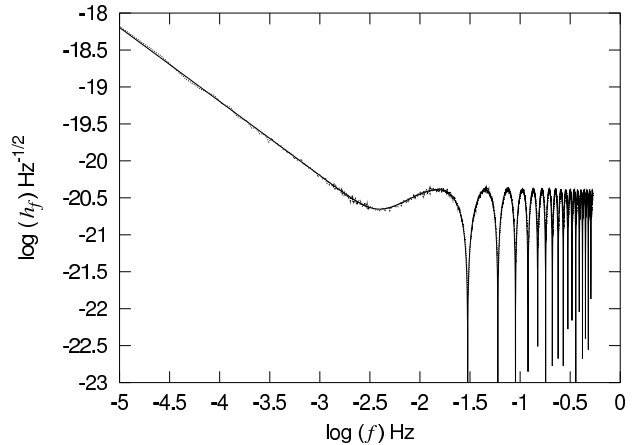}
\end{center}
\caption{A comparison of the simulated rms $X$ noise (dotted) to the
  analytical prediction (solid).}
\label{XnoiseComp}
\end{figure}

To arrive at a simulation of the $X$ noise we combine the noise
elements as dictated by Eq.~(\ref{xsig}). Doing so gives the results
displayed in Fig.~\ref{Xnoise}, which agrees with the predicted
results.
To see this, we start with the analytical expression of the average
Michelson noise curve shown in Fig.~\ref{Mnoise},
\begin{equation}
h_f^M(f) = \frac{1}{2L} \left( 4 S_{\rm ps} + 8 \left(1 + \cos^2
  \left( \frac{f}{f_\ast} \right) \right) \frac{S_{\rm acc}}{(2\pi
  f)^4} \right)^{1/2}
\end{equation}
which is derived in the appendix of Ref.~\cite{C01}.  In the above
$f_\ast = 1/(2\pi L)$ is the mean transfer frequency for an arm.
Next, we note that the $X$ signal is formed be differencing two
Michelson signals, one time delayed by roughly twice the light travel
time between spacecraft.  Therefore, the noise will enter in the $X$
signal as
\begin{equation}
n_X(t) = n_M(t) - n_M(t-2L) \,,
\end{equation}
which has a Fourier transform of
\begin{equation}
\widetilde{n}_X(f) = \widetilde{n}_M(f) \left( 1 - e^{-2i f / f_\ast}
\right)
\end{equation}
and a power spectral density of
\begin{equation}
S_X(f) = 4 \sin^2 \left(\frac{f}{f_\ast} \right) S_M(f) \,.
\end{equation}
The strain spectral density of the $X$ noise is given by
\begin{eqnarray} \label{eq:hfX}
h_f^X(f) &=& \sqrt{S_X(f)} \nonumber\\
&=& 2 \left\vert \sin\left( \frac{f}{f_\ast} \right) \right\vert 
h_f^M(f) \,.
\end{eqnarray}
Shown in Fig. \ref{XnoiseComp} is a plot of $h_f^X(f)$ along with the
average from Fig. \ref{Xnoise}.
Although the derivation of the $X$ noise strain spectral density assumed
constant arm-lengths we see that there is excellent agreement between
the predicted results of Eq.~(\ref{eq:hfX}) and the simulation, which
included the variations in the arms.

Although Eqs.~(\ref{dell}), (\ref{phi}), and (\ref{xsig}) give the
full response of a space-borne detector, they are analytically
difficult to handle and time consuming to evaluate. For this reason we
will now explore some approximations to the full response that use
information about the input waveforms and a simplified description of
the detector. These approximations not only aid in the development of
data analysis techniques, but also give a greater insight into the
workings of the detector.

\section{Low frequency approximation} \label{Slfa}

In sections \ref{Ssbd} and \ref{Sdra} we saw that the full response of
a space-borne gravitational wave detector was complicated by the
intrinsic arm-length fluctuations, pointing ahead, and the
signal-cancellation accounted for in the transfer functions.  As a
first approximation to the response of LISA we will neglect all of
these effects.  That is, we will work to linear order in the
spacecraft positions, evaluate all spacecraft locations at a common
time, and set the transfer functions to unity.  It should be noted
that this approximation was originally worked out by Cutler \cite{C98}
and can be viewed as an extension of the LIGO response to space-borne
detectors.  The transfer function ${\cal T}(f,t,\hat{k})$ can be set
equal to unity when $f \ll f_\ast$. For the LISA mission,
whose bandwidth is $10^{-5}$ to 1 Hz, the transfer frequency has a
mean value of $f_\ast = 0.00954 \approx 10^{-2}$ Hz.

In the limit $f \ll f_\ast$ and $f / \dot f \ll L$ the path length variation
(\ref{dell}) reduces to
\begin{eqnarray}\label{low}
\delta \ell_{ij}(t) &\simeq& \frac{1}{2} \, \frac{\hat{r}_{ij}(t)
  \otimes \hat{r}_{ij}(t)}{1-\hat{k} \cdot \hat{r}_{ij}(t)} :{\bf
  h}(\xi(t)) \left(\xi_j -\xi_i \right) \nonumber \\
&=& L \left[\hat{r}_{ij}(t) \otimes \hat{r}_{ij}(t):{\bf h}(\xi(t))
\right]\,.
\end{eqnarray}
Working in terms of strains and neglecting noise, the Michelson
signal from spacecraft 1 is given by
\begin{eqnarray}
s_1(t) &=& \frac{\delta\ell_{12}(t-2 L) + \delta\ell_{21}(t-L)}{2 L}
\nonumber\\
&& - \frac{\delta\ell_{13}(t-2 L) + \delta\ell_{31}(t-L)}{2 L} \nonumber \\
&\simeq& \frac{\delta\ell_{12}(t) + \delta\ell_{21}(t)
-\delta\ell_{13}(t) - \delta\ell_{31}(t)}{2 L} \, .
\end{eqnarray}
The last line follows from the condition $f \ll f_\ast$.

Using (\ref{wave}), (\ref{harmony}), and (\ref{low}) the strain can be
re-expressed as
\begin{eqnarray} \label{lfstrain}
s_1(t) = h_+(\xi_1(t))F^+ + h_\times(\xi_1(t)) F^\times \,,
\end{eqnarray}
where 
\begin{eqnarray}
\xi_1(t) &=& t-\hat{k}\cdot{\bf x}_{1}(t) \nonumber \\
&=& t - R \sin \theta \cos\big(\alpha(t) -\phi\big) 
\end{eqnarray}
is the gravitational wave phase measured at spacecraft 1.  The antenna
beam pattern factors, $F^+(t)$ and $F^\times(t)$, are given by
\begin{eqnarray}
F^+(t) &=& \frac{1}{2} \Big( \cos(2\psi) D^+(t) - \sin(2\psi)
D^\times(t) \Big) \nonumber\\
F^\times(t) &=& \frac{1}{2} \Big( \sin(2\psi) D^+(t) + \cos(2\psi)
D^\times(t) \Big) \, ,
\end{eqnarray}
where
\begin{eqnarray}
D^+(t) &=& \Big( \hat{r}_{12}(t) \otimes \hat{r}_{12}(t) -
  \hat{r}_{13}(t) \otimes \hat{r}_{13}(t) \Big) : {\bf e}^+
  \nonumber\\
D^\times(t) &=& \Big( \hat{r}_{12}(t) \otimes \hat{r}_{12}(t) -
  \hat{r}_{13}(t) \otimes \hat{r}_{13}(t) \Big) : {\bf e}^\times .
\end{eqnarray}
Working to linear order in the eccentricity, the Keplerian orbits given
in (\ref{keporb}) yield
\begin{eqnarray}
D^+(t) &=& \frac{\sqrt{3}}{64} \Bigg[ -36\sin^2 (\theta)
    \sin\big(2\alpha(t) - 2\lambda \big) \nonumber\\
&& + \big(3 + \cos(2\theta)\big) \bigg( \cos(2\phi) \Big(9
    \sin(2\lambda) \nonumber\\
&& - \sin\big( 4\alpha(t) - 2\lambda\big)\Big) + \sin(2\phi)
    \nonumber\\
&& \Big(\cos\big(4\alpha(t) - 2\lambda\big) - 9 \cos(2\lambda) \Big)
    \bigg) \nonumber\\
&& - 4\sqrt{3} \sin(2\theta) \Big (\sin\big(3\alpha(t) - 2\lambda -
    \phi\big) \nonumber\\
&& - 3 \sin\big(\alpha(t) - 2\lambda + \phi\big) \Big) \Bigg]
\end{eqnarray}
and
\begin{eqnarray} \label{dtimes}
D^\times(t) &=& \frac{1}{16} \Bigg[ \sqrt{3} \cos(\theta) \Big( 9
  \cos\big(2\lambda - 2\phi\big) - \cos\big(4 \alpha(t) \nonumber\\
&& - 2\lambda - 2\phi\big) \Big) - 6\sin(\theta)
  \Big(\cos\big(3\alpha(t) - 2\lambda - \phi \big) \nonumber\\
&& + 3\cos\big(\alpha(t) - 2\lambda + \phi\big) \Big) \Bigg] \, .
\end{eqnarray}
Equations (\ref{lfstrain}) to (\ref{dtimes}) constitute the analytical
formalism for the {\it Low Frequency Approximation}.  These equations
are numerically quick to evaluate and can be handled analytically.  As
a point of reference, the strain presented in Eq. (\ref{lfstrain}) can
be shown to be equivalent (most easily through a numerical comparison)
to that derived by Cutler \cite{C98}.

\begin{figure}[!tb]
\begin{center}
\includegraphics[width=0.48\textwidth]{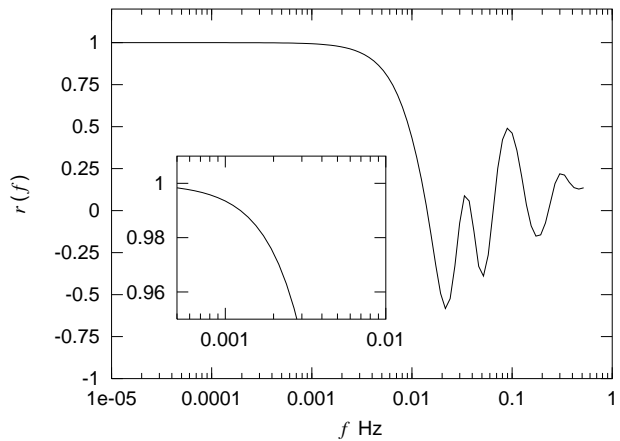}
\caption{The correlation between the {\it Low Frequency
  Approximation} and the full response of the LISA detector for a
  monochromatic source.  The oscillatory structure at high frequencies
  is due to the transfer functions introduced by the full response.}
\label{LowSimCor}
\end{center}
\end{figure}

To test the range of validity of this approximation we used {\it The
LISA Simulator} (TLS) as a template to calculate the correlation
between the full response and the {\it Low Frequency Approximation}
(LFA),
\begin{equation}
r(f) = \frac{\langle s_{\rm TLS} | s_{\rm LFA}\rangle} {\sqrt{ \langle
s_{\rm TLS}^2\rangle \langle s_{\rm LFA}^2\rangle}} \,.
\end{equation}
Using fixed random choices for the source location and orientation we
systematically varied the gravitational wave frequency and calculated
the correlation at each frequency.  The results of this calculation
are shown in Fig. \ref{LowSimCor}.

We found that the {\it Low Frequency Approximation} has a
strong correlation to the true response for frequencies below 3 mHz,
at which point the correlation drops to 95\%.  The steep turn down in
the correlation as the transfer frequency is approached is to be
expected as the {\it Low Frequency Approximation} neglects the
self-cancellation effects encoded in the transfer functions.  The
wiggles at higher frequencies are due to the transfer functions
present in the full response template $s_{\rm TLS}$.  The precise
structure of these oscillations depends on the source location through
the $\hat{k} \cdot \hat{r}_{ij}(t)$ dependence in the transfer
functions.  However, the turn down at 3 mHz is location independent.
The location dependence does not become strongly evident until the
correlation value has dropped to roughly zero.

The significance of a particular correlation value is dependent on the
signal-to-noise ratio of the source.  For high S/N the effects
neglected in the approximation will be detectable.  Conversely, for a
low S/N one may continue to use the approximation at higher
frequencies as the difference would not be noticeable.

\section{Rigid adiabatic approximation} \label{Sraa}

\subsection{Response formalism}

The breakdown of the {\it Low Frequency Approximation} comes about
through neglecting the transfer functions.  As a second approximation
to the LISA response we will now include the transfer functions, but
continue to hold the detector rigid by working to leading order in the
spacecraft positions and evaluating all spacecraft locations at the
same instant of time. Such an approximation has been worked out before
for the case of a stationary detector in \cite{CL01, C01}, but here we
extend it to include the motion of the detector.

Physically this approximation can be viewed in the following way.  At
an instant of time we hold the detector fixed and send photons up and
back along the interferometer arms and calculate the phase difference.
We then increment the time by a small amount, moving the rigid
detector to its new position in space, and repeat the process.  This
sequence of stationary states is the origin of the term ``Adiabatic''
for describing the approximation.

For chirping sources the Adiabatic approximation requires that the
frequency evolution $\dot f$ occurs on a timescale long compared to
the light travel time in the interferometer: $f / \dot f \ll L$.
When this condition does not hold the {\it Rigid Adiabatic
Approximation} is no longer valid and the full response should be
used. In the limit $f / \dot f \ll L$ the path length variation (\ref{dell})
reduces to
\begin{equation}\label{delRA}
\delta\ell_{ij}(\xi) = L \sum_n {\bf
  D}(f_n,t,\hat{k}):{\bf h}_n(\xi)\, ,
\end{equation}
where the one-arm detector tensor is given by
\begin{equation}
{\bf D}(f,t,\hat{k}) = \frac{1}{2}\Big(\hat{r}_{ij}(t) \otimes
  \hat{r}_{ij}(t)\Big) {\mathcal T}(f,t,\hat{k}) \, ,
\end{equation}
and the transfer function is
\begin{eqnarray}
{\mathcal T}(f,t,\hat{k}) &=& \textrm{sinc} \left( \frac{f}{2
f_*}\Big(1 - \hat{k} \cdot \hat{r}_{ij}(t)\Big) \right) \nonumber\\
&& \times \exp \left( i \frac{f}{2 f_*}\Big(1 - \hat{k} \cdot
    \hat{r}_{ij}(t) \Big) \right) \, .
\end{eqnarray}
The Michelson signal is given by
\begin{eqnarray}
s_1(t) &=& \frac{\delta\ell_{12}(t-2 L) + \delta\ell_{21}(t-L)}{2 L}
\nonumber\\
&& - \frac{\delta\ell_{13}(t-2 L) + \delta\ell_{31}(t-L)}{2 L} \,,
\end{eqnarray}
which may now be expressed as
\begin{equation} \label{RAstrain}
s_1(t) = \sum_n {\bf D}(\hat{k},f_n):{\bf h}_n(\xi) \,,
\end{equation}
and the round-trip detector tensor takes the form
\begin{equation}
{\bf D}(\hat{k},f) = \frac{1}{2} \Big((\hat{a} \otimes \hat{a})
{\mathcal T}(\hat{a} \cdot \hat{k},f) - (\hat{b} \otimes \hat{b})
{\mathcal T}(\hat{b} \cdot \hat{k},f)\Big) \,,
\end{equation}
and the round-trip transfer function is
\begin{eqnarray}
{\mathcal T}({\bf a} \cdot \hat{k},f) &=& \frac{1}{2} \left[ 
  \textrm{sinc} \left( \frac{f(1-\hat{a}\cdot\hat{k})}{2f_\ast} 
  \right) \right. \nonumber\\
&& \times \exp \left( -i \frac{f}{2f_\ast} (3+\hat{a}\cdot\hat{k})
  \right) \nonumber\\
&& + \textrm{sinc} \left( \frac{f(1+\hat{a}\cdot\hat{k})}{2f_\ast}
  \right) \nonumber\\
&& \left. \times \exp \left( -i \frac{f}{2f_\ast} 
  (1+\hat{a}\cdot\hat{k}) \right) \right] \,.
\end{eqnarray}
The time dependent unit vectors, $\hat{a}(t)$ and $\hat{b}(t)$, are
given by
\begin{eqnarray} \label{eq:abuv}
\hat{a}(t) &=& \frac{{\bf x}_2(t) - {\bf x}_1(t)}{L} \nonumber\\
\hat{b}(t) &=& \frac{{\bf x}_3(t) - {\bf x}_1(t)}{L} \,.
\end{eqnarray}
Collectively these equations are the analytical formalism for the {\it
Rigid Adiabatic Approximation}. As with the {\em Low Frequency
Approximation}, the expressions are computationally quick to evaluate
and can be easily manipulated analytically.

\begin{figure}[!tb]
\begin{center}
\includegraphics[width=0.48\textwidth]{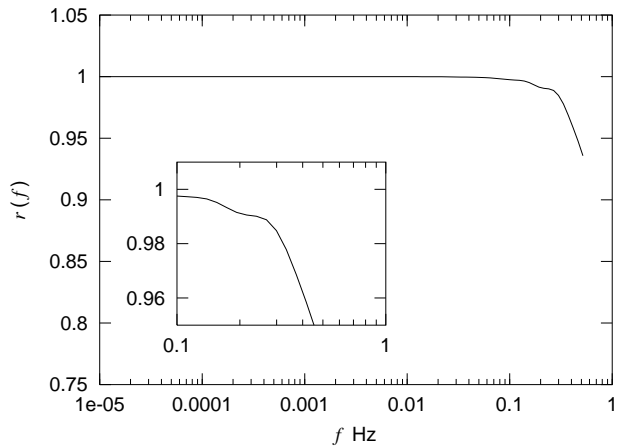}
\end{center}
\caption{The correlation between the {\it Rigid Adiabatic
  Approximation} and the full response of the LISA detector.  The turn
  down at $\sim$500 mHz is due to neglecting the higher order
  effects in the spacecraft positions.}
\label{fRAcor}
\end{figure}

Figure \ref{fRAcor} shows the correlation between the full response
and the {\it Rigid Adiabatic Approximation} for a monochromatic
gravitational wave.  Note that by including the transfer functions we
are able to extend agreement with the full response two decades in
frequency beyond where the {\it Low Frequency Approximation} broke
down.
The turn down at $\sim 0.5$ Hz comes about through neglecting the
second order terms in the spacecraft positions. As we described in
Sec.~\ref{Ssbd-oe} the second order orbital effects are down by two
orders of magnitude in comparison to the linear order. This shows up
in the {\it Rigid Adiabatic Approximation} through the transfer
frequencies, which are evaluated for a rigid detector. Normally the
transfer frequencies are given by
\begin{equation}
f^*_{ij}(t) = \frac{1}{2\pi \ell_{ij}(t)} \,,
\end{equation}
but for a rigid detector this reduces to the static form $f_\ast =
1/(2\pi L)$. The extension to higher orders in the orbital
eccentricity can be done. The trade off is that the expressions become
more complicated since the transfer frequencies would then become
functions of time.  In turn, this would require that each transfer
frequency be evaluated along each arm during each time step rather
than using one constant value throughout the entire calculation.
Additionally, the normalization of the unit vectors in
Eq.~(\ref{eq:abuv}) would need to be evaluated at each step since the
arm-lengths would vary as a function of time via Eq.~(\ref{ovar}).
Such an approach would be appropriately called the {\it Flexing
Adiabatic Approximation} since the arm-lengths would now oscillate in
time about a mean value of $L$.  Although the expressions would become
analytically complicated, the numerical evaluation would not be
significantly slower since the additional steps are straightforward to
evaluate.
\begin{figure}[!tb]
\begin{center}
\includegraphics[width=0.48\textwidth]{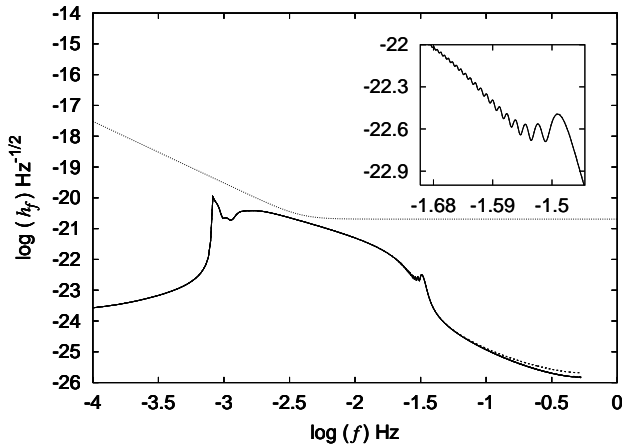}
\end{center}
\caption{A comparison of strain spectral densities as calculated
using the \emph{Rigid Adiabatic Approximation} (dashed) and the full
response (solid) for a $5000 M_\odot$ intermediate mass black hole
binary at $z=1$. For reference, the dotted line is the average
Michelson noise in the detector.}
\label{BH_BH_Comp}
\end{figure}

Figure \ref{BH_BH_Comp} compares the output of {\it The LISA Simulator} to the
{\it Rigid Adiabatic Approximation} for a binary system of intermediate
mass black holes, each of mass $5000 M_\odot$ at a redshift of $z=1$. The
observation covers the final year before coalescence. The agreement is
excellent.

In a recent paper \cite{seto}, Seto used a variant of the {\it Rigid
Adiabatic Approximation} to calculate the effects of LISA's finite
arm-lengths on the analysis of gravitational waves from chirping
supermassive binary black holes. A comparison of
our {\it Rigid Adiabatic Approximation}, which is derived from the
path length variation, and Seto's approach, which is based
on the Doppler shift formalism is given in Appendix C.

\subsection{Applications}

Utilizing the speed of the {\it Rigid Adiabatic Approximation} we
may investigate various data analysis questions. Here we provide one
concrete example by determining when phase evolution of a binary
system due to radiation reaction needs to be included in the source
modeling.

For our calculations we used the restricted
Post-Newtonian approximation, whereby the gravitational wave amplitude
is calculated to first order, while the phase evolution is calculated
to second order~\cite{BIWW96}. The justification for this is that
LISA will be far more sensitive to the phase than the
amplitude~\cite{C98}. The lack of additional harmonics of the orbital
period also simplifies the calculation as we only have to calculate a
single transfer function at each time step.

\begin{figure}[!tb]
\begin{center}
\includegraphics[width=0.48\textwidth]{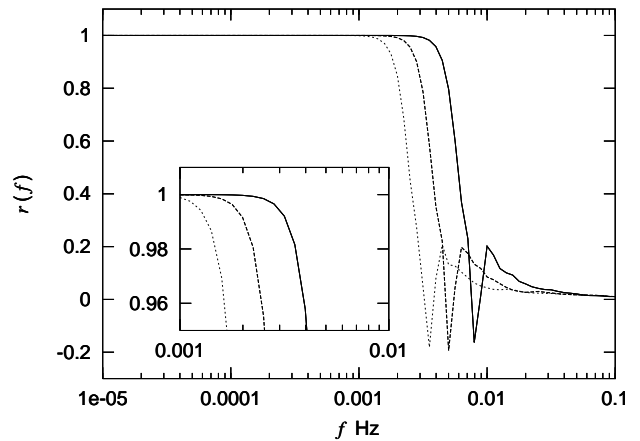}
\end{center}
\caption{The correlation between a monochromatic {\it Rigid Adiabatic
  Approximation} and one that includes a 2PN phase evolution.  The
  solid line represents a WD-WD binary with mass components of 0.5
  $\textrm{M}_{\odot}$, the dashed line is for a NS-NS binary with
  masses 1.4 $\textrm{M}_{\odot}$, and the dotted line is for a 10
  $\textrm{M}_{\odot}$ black hole and a 1.4 $\textrm{M}_{\odot}$
  neutron star binary.}
\label{RA-RA2PN_Freq}
\end{figure}

To quantify the importance of including the evolution of the
gravitational wave phase, we calculated the correlation between a
monochromatic {\it Rigid Adiabatic Approximation} to one in which the
phase evolution is included.  Figure~\ref{RA-RA2PN_Freq} shows the
correlation for three types of binaries expected to reside inside our
own galaxy: a white dwarf binary with mass components 0.5
$\textrm{M}_{\odot}$, a neutron star binary with masses 1.4
$\textrm{M}_{\odot}$, and a 10 $\textrm{M}_{\odot}$ black hole with a
neutron star companion.

What we found is that the frequency at which the monochromatic signal
diverges from one that includes phase evolution depends on the masses
of the binary components.  The reason for this comes from the
expression for $\dot{f}$, which contains a mass dependent coefficient.
For the stellar mass binaries we studied, the drop in the correlation
happened to coincide with the breakdown of the {\it Low Frequency
Approximation}. Thus, most Milky Way sources can be modeled as
monochromatic sources using the {\it Low Frequency Approximation}.

\begin{figure}[!tb]
\begin{center}
\includegraphics[width=0.48\textwidth]{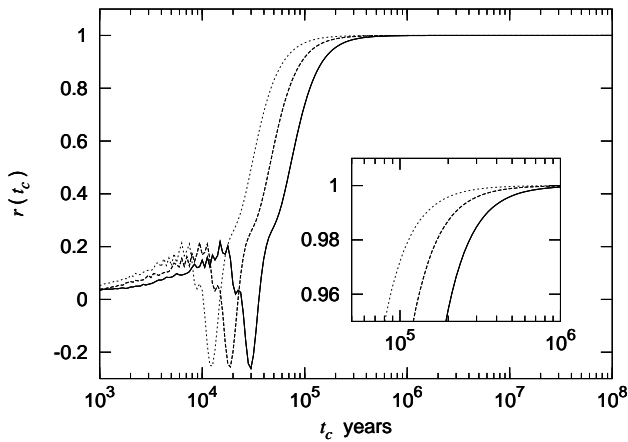}
\end{center}
\caption{The correlation between a monochromatic {\it Rigid
  Adiabatic Approximation} and one that includes a 2PN phase evolution,
  this time expressed in terms of the time to coalescence.}
\label{RA-RA2PN_CTime}
\end{figure}

Another way to represent the same data
is to map the initial frequency to the time of coalescence.  The
results of this calculation are shown in Fig.~\ref{RA-RA2PN_CTime}.
In this case we see that chirping becomes important for stellar mass
sources within $\sim$$10^5$ years of coalescence.  As expected, the
mapping to the new variable preserves the mass dependence seen in
Fig.~\ref{RA-RA2PN_Freq}.

\begin{figure}[!tb]
\begin{center}
\includegraphics[width=0.48\textwidth]{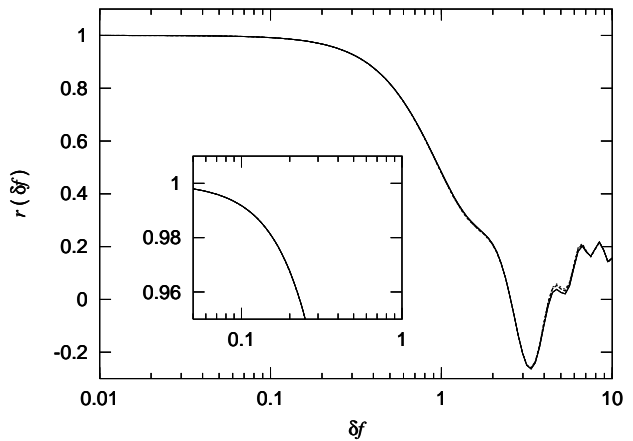}
\end{center}
\caption{The correlation as a function of the fractional bin width
  change in the frequency, $\delta f = (f_f - f_i) / \Delta f$,
  between a monochromatic {\it Rigid Adiabatic Approximation} and one
  that includes a 2PN phase evolution. Note that all three types of
  sources previously considered are included in this plot.}
\label{RA-RA2PN_df}
\end{figure}

A final way to represent this data is to set the independent variable
equal to the change in the frequency scaled by a bin width, $\delta f =
(f_f - f_i) / \Delta f$, where for one year of observation the bin
width is $\Delta f = 1/\textrm{yr} \simeq 3.2 \times 10^{-8}$ Hz. Such an
approach is shown in Fig.~\ref{RA-RA2PN_df}. Unlike with the other
representations of the correlation between a monochromatic and
coalescing signal, the results of this calculation are independent of
the system's masses.  It is also interesting to note that this
result implies that it will be possible to detect if a source is
coalescing or not well within a bin width.  This fact is not in
conflict with the Nyquist theorem, which states that the frequency
resolution will not be better than the inverse of the observation
time.  The reason being that we have additional information, namely
the functional form of the phase evolution, which is not assumed in
deriving Nyquist's theorem.

\section{Discussion} \label{Sdis}

We have examined the forward modelling of space-borne gravitational
wave detectors with special emphasis on the LISA observatory.
Forward modelling will play two distinct roles in the developement
of space-borne observatories. The first is as part of a complete
end-to-end model that takes into account every concievable physical
effect, and the second is as an intermdeiary between source simulation
and data analysis. Here we have focussed on the latter role, and to
that end we have studied two simple approximations
to the full response - the {\it Low Frequency Approximation}
and the {\it Rigid Adiabatic Approximation}. We found that
the {\it Rigid Adiabatic Approximation} could be used in place of
the full response for a wide range of data analysis projects. For example, the
relatively simple analytic form of the {\it Rigid Adiabatic Approximation}
is well suited to the calculation of Fisher information matrices
in studies of astrophysical parameter extraction. On the other hand,
{\it The LISA Simulator} is available if we need to simulate the
response to highly relativistic gravitational wave sources such as the merger
of two black holes.

\begin{acknowledgments}
We would like to thank Ron Hellings for a number of helpful
discussions concerning the simulation of the LISA noise and the
removal of orbital effects.
\end{acknowledgments}

\appendix

\section{Spacecraft Positions}

For a constellation of spacecraft in individual Keplerian orbits with
an inclination of $i = \sqrt{3} e$ the coordinates of each spacecraft
are given by the expressions
\begin{eqnarray} \label{TIpos}
x &=& r \left( \cos(\sqrt{3}e)\cos\beta\cos\gamma -
\sin\beta\sin\gamma \right) \nonumber\\
y &=& r \left( \cos(\sqrt{3}e)\sin\beta\cos\gamma +
\cos\beta\sin\gamma \right) \nonumber\\
z &=& -r \sin(\sqrt{3}e)\cos\gamma
\end{eqnarray}
where $\beta = 2n\pi /3 + \lambda$ ($n=0,1,2$) is the relative orbital
phase of each spacecraft in the constellation, $\gamma$ is the
ecliptic longitude, and $r$ is the standard Keplerian radius
\begin{equation}
r = \frac{R(1-e^2)}{1+e\cos\gamma} \, .
\end{equation}
Here $R$ is the semi-major axis of the guiding center and has an
approximate value of 1 AU.
\begin{figure}[!tb]
\begin{center}
\input{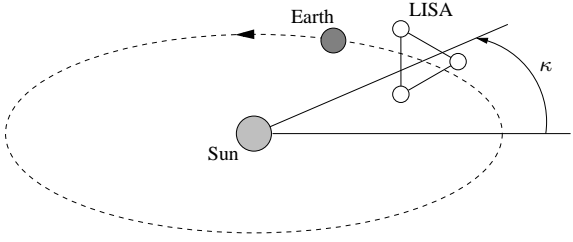}
\end{center}
\caption{The angle $\kappa$ gives the initial ecliptic longitude of
  the guiding center.}
\end{figure}
\begin{figure}[!tb]
\begin{center}
\input{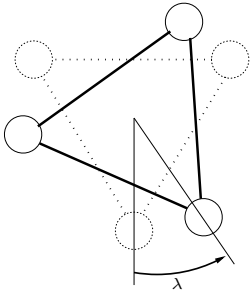}
\end{center}
\caption{As viewed by an observer at the origin, $\lambda$ gives the
  initial orientation of the spacecraft constellation.}
\end{figure}

To get the above coordinates as a function of time we first note that
the ecliptic longitude is related to the eccentric anomaly, $\psi$, by
\begin{equation} \label{az-ea}
\tan\left(\frac{\gamma}{2}\right) =
\sqrt{\frac{1+e}{1-e}}\tan\left(\frac{\psi}{2}\right) \, ,
\end{equation}
and the eccentric anomaly is related to the orbital phase $\alpha(t) =
2\pi t / T + \kappa$ via Kepler's equation
\begin{equation} \label{Kepler_Eq}
\alpha - \beta = \psi - e\sin\psi \, .
\end{equation}
Assuming a small eccentricity we may solve Eq. (\ref{Kepler_Eq})
through an iterative process where we treat the $e\sin\psi$ term as
being lower order than $\psi$,
\begin{equation}
\psi_n = \alpha-\beta + e\sin\psi_{n-1} \; .
\end{equation}
Through such a procedure we arrive at
\begin{equation}
\psi = \alpha-\beta + e\sin(\alpha-\beta) +
e^2\cos(\alpha-\beta)\sin(\alpha-\beta) + \cdots \; .
\end{equation}
Substituting this result into Eq. (\ref{az-ea}) and expanding to
second order in the eccentricity gives an ecliptic longitude of
\begin{equation}
\gamma = (\alpha-\beta) + 2 e \sin(\alpha-\beta) + \frac{5}{2} e^2
\cos(\alpha-\beta)\sin(\alpha-\beta) + \cdots \,.
\end{equation}
Substituting the ecliptic longitude series into Eq. (\ref{TIpos}) and
keeping terms up to order $e^2$ gives the Cartesian positions of the
spacecraft as functions of time,
\begin{eqnarray}
x(t) &=& R \cos(\alpha) + \frac{1}{2} R e \Big( \cos(2\alpha-\beta) -
3\cos(\beta) \Big) \nonumber\\
&& + \frac{1}{8} R e^2 \Big( 3\cos(3\alpha-2\beta) - 10\cos(\alpha) 
\nonumber\\
&& - 5\cos(\alpha-2\beta) \Big) \nonumber\\
y(t) &=& R \sin(\alpha) + \frac{1}{2} R e \Big( \sin(2\alpha-\beta) -
3\sin(\beta) \Big) \nonumber\\
&& + \frac{1}{8} R e^2 \Big( 3\sin(3\alpha-2\beta) - 10\sin(\alpha)
 \nonumber\\
&& + 5\sin(\alpha-2\beta) \Big) \nonumber\\
z(t) &=& -\sqrt{3} R e \cos(\alpha-\beta) \nonumber\\
&&+ \sqrt{3} R e^2 \Big( \cos^2(\alpha-\beta) 
+ 2\sin^2(\alpha-\beta) \Big) \; .
\end{eqnarray}
These are the desired coordinates of each spacecraft as a function of
time.

\section{Doppler style derivation of the full response}

The Doppler shift of a photon emitted by spacecraft 1 and received by
spacecraft 2 can be elegantly derived~\cite{EW75} using the symmetries of the
spacetime (\ref{combo}). When $\phi=0$ the spacetime admits three
Killing vectors,
\begin{equation}\label{kv}
\vec\zeta_{(1)} = \hat{u}, \quad \vec\zeta_{(2)} = \hat{v}, \quad 
\vec\zeta_{(3)} = \vec{\partial_t}+\hat{k} \, .
\end{equation}
These provide three constants of the motion, $\vec\zeta_{(i)}\cdot
\vec{U}$, which along with the normalization condition
$\vec{U}\cdot\vec{U}=0$ or $\vec{U}\cdot\vec{U}=-1$, fully specify
$\vec{U}(\lambda)$ in terms of some initial four velocity
$\vec{U}(0)$. Writing the metric as
$g_{\mu\nu}=\eta_{\mu\nu}+h_{\mu\nu}$, we may express the photon
propagation four-vector as
\begin{equation}
\sigma^\mu = s^\mu - \frac{1}{2}
\eta^{\mu\alpha}h_{\alpha\beta}s^\beta \, ,
\end{equation}
where $\vec{s}$ is a null vector in the unperturbed geometry: $s^\mu
s^\nu \eta_{\mu\nu}=0$.  At the time of emission from spacecraft 1,
$\vec{s}(t_1)=\vec{s}_0+\delta\vec{s}_1$, while at the time of
reception at spacecraft 2,
$\vec{s}(t_2)=\vec{s}_0+\delta\vec{s}_2$. Here $\vec{s}_0$ is parallel
to the unit vector connecting the two spacecraft in the unperturbed
spacetime, while $\delta\vec{s}_1$ and $\delta\vec{s}_2$ are
perturbations to the path due to lensing by the gravitational
wave. Defining
\begin{eqnarray}\label{deltas}
&&\Delta s^\alpha = s^\alpha(t_2)- s^\alpha(t_1), \nonumber \\
&&\Delta h_{\alpha\beta} = h_{\alpha\beta}(t_2)-h_{\alpha\beta}(t_1),
\end{eqnarray}
we have
\begin{eqnarray}
&& \sigma^\alpha(t_2)=\sigma^\alpha(t_1)+\Delta s^\alpha - \frac{1}{2}
\eta^{\alpha\beta} \Delta h_{\beta\gamma} s^\gamma \nonumber \\
&& g_{\alpha\beta}(t_2) =g_{\alpha\beta}(t_1)+\Delta h_{\alpha\beta}
\end{eqnarray}
which yields
\begin{eqnarray}\label{null}
2 s^\alpha \Delta s^\beta \eta_{\alpha\beta}&=&
\sigma^\alpha(t_2)\sigma^\beta(t_2)g_{\alpha\beta}(t_2) \nonumber \\
&& - \sigma^\alpha(t_1)\sigma^\beta(t_1)g_{\alpha\beta}(t_1) \nonumber \\
&=& 0 \, .
\end{eqnarray}
Equations (\ref{kv}), (\ref{deltas}) and (\ref{null}) yield four
equations for the four $\Delta s^\alpha$:
\begin{eqnarray}\label{Ds}
&& \zeta^\alpha_{(i)}\Delta s^\beta \eta_{\alpha\beta} = -\frac{1}{2}
s^\alpha_0\zeta^\beta_{(i)} \Delta h_{\alpha\beta} \nonumber \\
&& s^\alpha_0 \Delta s^\beta \eta_{\alpha\beta} = 0 \, .
\end{eqnarray}
These can be solved to give, for example,
\begin{equation}
\Delta s^t = -\frac{ s^i_0 s^j_0 \Delta h_{ij}}{2 \vec{k}\cdot\vec{s}} \, .
\end{equation}
Here $\vec{k} \rightarrow (1,\hat{k})$ is the null propagation vector
for the gravitational wave. The frequencies of the emitted and
received photons, as measured at spacecraft 1 and spacecraft 2, are
given by $\nu_1 = -\vec{U}_1(t_1) \cdot \vec{\sigma}(t_1)$ and $\nu_2
= -\vec{U}_2(t_2) \cdot \vec{\sigma}(t_2)$ respectively. Here
$\vec{U}_1$ and $\vec{U}_2$ are the four-velocities of the two
spacecraft.  Note the $\nu_1=\nu_0$ is the operating frequency of the
laser on board spacecraft 1.  Evaluating $\nu = -\vec{U}\cdot
\vec{\sigma}$ yields
\begin{equation}\label{freq}
\nu = -\gamma(t)\left( s^t(t)+v^i_1(t)s^j(t)\eta_{ij} + \frac{1}{2}
h_{ij}(t)v^i(t)s_0^j \right)
\end{equation}
where $U^t=\gamma=dt/d\tau$ and the $v^i$ are the ordinary three
velocities of the spacecraft.  The spacecraft trajectories $\vec{U}$
may be expressed in terms of the unperturbed trajectories $\vec{U}_0$
according to
\begin{equation}\label{sc}
U^\alpha = U^\alpha_0 + \eta^{\alpha i}h_{ij}U^j_0 + A^\alpha \, ,
\end{equation}
where $A^\alpha$ are constants set by the initial conditions at some
time $t$.  Once the initial conditions for the spacecraft have been
set, equations (\ref{Ds}), (\ref{freq}) and (\ref{sc}) give the full
Doppler shift $\nu_2-\nu_1$ at any subsequent time. The expressions
simplify considerably if we drop terms of order $v^2$, $vh$ and
higher:
\begin{equation}
\Delta\nu = \nu_2-\nu_1 \simeq \Delta s^t =  
-\frac{ s^i_0 s^j_0 \Delta h_{ij}}{2 \vec{k}\cdot\vec{s}_0} \, .
\end{equation}
Converting this into a fractional frequency shift, $\Delta \nu/\nu_0$,
and using ${\bf s}_0=\nu_0 \, \hat{a}$, where $\hat{a}$ is the unit
vector connecting the two spacecraft in the background geometry, we
have
\begin{equation}\label{freqshift}
\frac{\Delta\nu}{\nu_0} = \frac{ \hat{a}\otimes\hat{a} : \Delta {\bf
h}}{2(1-\hat{k}\cdot\hat{a})} \, .
\end{equation}
Integrating the above expression with respect to time yields the time
delay described by Eq.~(\ref{dell}).

\section{Reconciliation Between Alternative Rigid Adibatic Formalisms}

According to Eq.~(2.2) of Ref.~\cite{seto2}, the relative frequency
shift for a photon traveling from spacecraft 2 to spacecraft 1 can be
expressed as
\begin{eqnarray} \label{eq:y}
y_{31}(t) &=& \frac{1}{2} \big( A_+ \cos(2\psi_{12}) + i A_\times
\sin(2\psi_{12}) \big) \nonumber\\
&& \times \big( 1 - \cos\theta_{12} \big) \big( U(t,1) - U(t-L,2)
\big) \,,
\end{eqnarray}
where the function $U(t,i)$ gives the phase of the gravitational
wave at spacecraft $i$, $\theta_{ij}$ is the angle
between the source location on the sky, $-\hat{k}$, and the detector
arm ${\bf x}_i - {\bf x}_j$,
\begin{equation} \label{eq:theta}
\cos \theta_{ij} = -\hat{k} \cdot \hat{r}_{ji} \,,
\end{equation}
and $\psi_{ij}$ is given through the relationship
\begin{equation} \label{eq:psi}
\tan \psi_{ij} = \frac{\hat{r}_{ji} \cdot \hat{q}}{\hat{r}_{ji} \cdot
  \hat{p}} \,.
\end{equation}
Here $\hat{p}$ and $\hat{q}$ are unit vectors along the principle
polarization axes of the gravitational wave.  The amplitude
coefficients, $A_+$ and $A_\times$, are given as functions of the
orbital inclination angle $\iota$ and the intrinsic amplitude $A$,
\begin{eqnarray} \label{eq:As}
A_+ &=& A (1+\cos^2\iota) \nonumber\\
A_\times &=& 2 A \cos\iota\,.
\end{eqnarray}
By direct substitution of Eqs.~(\ref{eq:theta}), (\ref{eq:psi}), and
(\ref{eq:As}) into the relative frequency shift we arrive at
\begin{eqnarray}
y_{31}(t) &=&  \frac{1 - (\hat{k} \cdot
  \hat{r}_{21})^2}{ \left(\hat{r}_{21} \cdot
  \hat{p}\right)^2 + \left(\hat{r}_{21} \cdot \hat{q}\right)^2}
  \nonumber\\
&& \times \Big( A_+ \big( (\hat{r}_{21} \cdot \hat{p})^2 -
(\hat{r}_{21} \cdot \hat{q})^2 \big) \nonumber\\
&& + 2 i A_\times (\hat{r}_{21} \cdot \hat{p})(\hat{r}_{21} \cdot
\hat{q}) \Big) \nonumber\\
&& \times \frac{\big( U(t,1) - U(t-L,2) \big)}
{2 (1-\hat{k} \cdot \hat{r}_{21})} \,.
\end{eqnarray}
The overall coefficient
\begin{equation}\label{coeff}
\frac{1-(\hat{k} \cdot \hat{r}_{21})^2}{ \left(\hat{r}_{21} \cdot
  \hat{p}\right)^2 + \left(\hat{r}_{21} \cdot \hat{q}\right)^2},
\end{equation}
can be shown to equal unity by writing
the vector $\hat{r}_{ij}$ in terms of the orthonormal triad
$\{\hat{p}, \hat{q}, \hat{k}\}$ according to
\begin{equation}
\hat{r}_{ij} = \sin \theta' \cos \phi' \hat{p} + \sin \theta' \sin
\phi' \hat{q} + \cos \theta' \hat{k} \,.
\end{equation}

To proceed further we note that the gravitational wave basis tensors
can be expressed as functions of $\hat{p}$ and $\hat{q}$,
\begin{eqnarray}
{\mbox{\boldmath$\epsilon$}}^+ &=& \hat{p} \otimes \hat{p} - \hat{q}
\otimes \hat{q} \nonumber\\
{\mbox{\boldmath$\epsilon$}}^\times &=& \hat{p} \otimes \hat{q} + \hat{q}
\otimes \hat{p} \,.
\end{eqnarray}
Using these relationships, it follows that 
\begin{eqnarray}
(\hat{r}_{ij} \otimes \hat{r}_{ij}) : {\mbox{\boldmath$\epsilon$}}^+
&=& (\hat{r}_{ij} \cdot \hat{p})^2 - (\hat{r}_{ij} \cdot \hat{q})^2 \\
(\hat{r}_{ij} \otimes \hat{r}_{ij}) :
{\mbox{\boldmath$\epsilon$}}^\times
&=& 2(\hat{r}_{ij} \cdot \hat{p})(\hat{r}_{ij} \cdot \hat{q}) \,.
\end{eqnarray}

Returning to the expression for the relative frequency shift we now have
\begin{eqnarray}\label{ourseto}
y_{31}(t) &=& \frac{\hat{r}_{21} \otimes
\hat{r}_{21}}{2(1-\hat{k} \cdot \hat{r}_{21})} : \big( A_+
{\mbox{\boldmath$\epsilon$}}^+ \nonumber + i A_\times
{\mbox{\boldmath$\epsilon$}}^\times \big) \nonumber\\ &&\times \big(
U(t,1) - U(t-L,2) \big) \nonumber\\
&=& \frac{\hat{r}_{21} \otimes
\hat{r}_{21}: \Delta {\bf h} }{2(1-\hat{k} \cdot \hat{r}_{21})} \,,
\end{eqnarray}
where in the last step we combined the amplitude and phase functions to
form the difference in the gravitational wave evaluated at each
spacecraft.  The above results agrees with Eq.~(\ref{freqshift})
found in Appendix B.

\end{document}